%% file: durban2011c.tex
\documentclass{kluwer}    

\usepackage{amsmath}

\usepackage{graphicx}

\newdisplay{guess}{Conjecture}

\input kigou.tex

\newcommand{\GA}{\alpha}
\newcommand{\GB}{\beta}
\newcommand{\GG}{\gamma}
\newcommand{\GD}{\delta}
\newcommand{\GE}{\epsilon}

\newcommand{\GC}{\psi}
\newcommand{\GX}{\chi}

\newcommand{\GP}{\phi}

\newcommand{\GJ}{\theta}

\newcommand{\pd}{\partial}

\newcommand{\be}{\begin{equation}}
\newcommand{\ee}{\end{equation}}

\begin{document}                                                                                   
\begin{opening}         
\title{Binary black hole circular orbits computed with {\sc cocal}} 
\author{Antonios Tsokaros}
\institute{Department of I.C.S.E., University of Aegean, Karlovassi 83200, Samos, Greece}
\author{K\=oji \surname{Ury\=u}}  
\institute{Department of Physics, University of the Ryukyus, Senbaru, Nishihara, 
Okinawa 903-0213, Japan}
\date{July 15, 2012}

\begin{abstract}
In this work we present our first results of binary black hole circular orbits using 
{\sc cocal}, the Compact Object CALculator. Using the 3+1 decomposition five equations are
being solved under the assumptions of conformal flatness and maximal slicing. Excision
is used and the appropriate apparent horizon boundary conditions are applied.
The orbital velocity is determined by imposing a Schwarzschild behaviour at infinity.
A sequence of equal mass black holes is obtained and its main physical characteristics
are calculated. 
\end{abstract}
\keywords{Black holes - Initial data - Computational methods}
\end{opening}

\section{Introduction}
\label{sec:int}

One of the most important tests of Einstein's theory of general relativity is the search
for gravitational waves. A great effort both in the experimental and theoretical problems has been
made and detection can happen almost any time. A highly probable scenario will be that the gravitational
wave is coming from a binary system of two black holes or two neutron stars or a black hole/neutron star
system. Therefore the extraction of a waveform that represents such configurations 
is an important step towards detection. 

From the mathematical point of view assuming spacetime is foliated by three dimensional hypersurfaces
$\Sigma_t$, Einstein's equations can be written as an initial value problem for the first and the second 
fundamental form of $\Sigma_t$. Then we get two sets of equations; one set that provides initial data
and another that evolves them to acquire the full spacetime. In \cite{COCAL} we provided a method to
solve the former set of equations and here we elaborate on these solutions and identify those that represent
circular orbits. Also we present some preliminary results regarding the physical characteristics of
these solutions.  

The spacetime metric on $\Sigma_t$ is written in 3+1 form as
\be
ds^{2} = g_{\mu\nu}dx^{\mu}dx^{\nu} = -\GA^{2}dt^{2}+\GG_{ij} (dx^{i}+\GB^{i}dt) (dx^{j}+\GB^{j}dt). 
\ee
We assume the spatial three metric $\GG_{ij}$ on the slice $\Sigma_t$ 
to be conformally flat $\GG_{ij} = \GC^4 f_{ij}$. Then
the system to be solved, which are Hamiltonian and momentum 
constraints and the spatial trace of the Einstein's equation, becomes
\begin{subequations}\label{eq:all}
\begin{align}
\Lap \GC &= -\frac{\GC^5}{8}{\tilde A}_{ij}{\tilde A}^{ij},  
\label{eq:HC}\\
\Lap \GB_i &= -2\,\alpha\, {\tilde A}_i{}^j \pd_j \ln\frac{\GC^6}{\GA}-\frac{1}{3}\pd_i\pd_j{\tilde\GB}^j,
\label{eq:MC}\\
\Lap (\GA\GC) &= \frac{7}{8}\,\alpha\,\GC^5{\tilde A}_{ij}{\tilde A}^{ij},  
\label{eq:Kdot}
\end{align}
\end{subequations}
where $\Lap:=\pa_i\pa^i$ is a flat Laplacian.
The field variables $\psi, \alpha$, and $\beta^i$ are the conformal factor, 
lapse, and shift vector, respectively.  We also assume maximal slicing 
to $\Sigma_t$, so that the trace $K$ of the extrinsic curvature 
$K_{ij}=A_{ij}+\frac{1}{3}\GG_{ij}K$ vanishes.  
The conformally rescaled quantity $\tilde A_{ij}$ becomes 
\beq
{\tilde A}_{ij} = 
\frac1{2\alpha} \left(\pd_i{\tilde \GB}_j + \pd_j{\tilde \GB}_i 
- \frac{2}{3}f_{ij}\pd_k{\tilde \GB}^k\right)\ ,
\eeq
where the derivative $\pa_i$ is associated with the flat metric 
$f_{ij}$, and conformally rescaled quantities with tilde are defined by 
$\tilde A_i{}^j= A_i{}^j$ and ${\tilde \beta}^i=\beta^i$, whose indexes 
are lowered (raised) by $f_{ij}$ ($f^{ij}$).

\section{Overview of the algorithm}
\label{sec:poisol}

\begin{figure}
\begin{center}
\includegraphics[height=70mm]{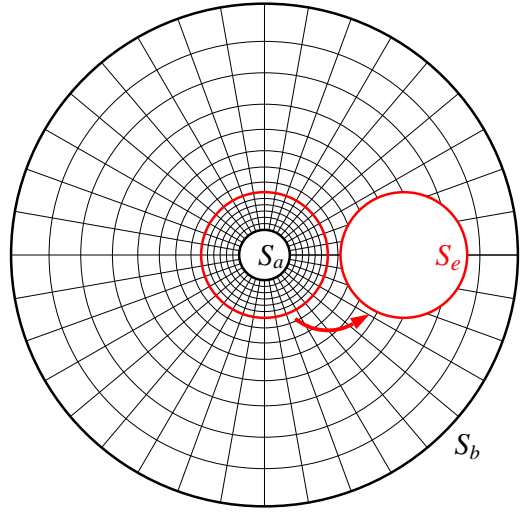}
\caption{The core {\sc cocal} coordinate system where the initial value equations are being
solved. One dimension is supressed. 
The sphere at the center, surface $S_a$, corresponds to one compact object. 
The radius of coordinate patch doesn't reflect the actual size.}
\label{fig:singlepatch}  
\end{center}
\end{figure}

In our previous paper \cite{COCAL}, we presented the new code for computing equilibriums 
of astrophysical compact objects -- {\sc cocal}, Compact Object CALculator.
In the {\sc cocal} code we cover the initial hypersurface $\Sigma_t$ with spherical grids like
the one that appears in Fig.\ref{fig:singlepatch}. Characteristic features are the inner spherical surface
$S_a$ of the patch with radius $r_a$ (small circle at the center of Fig.\ref{fig:singlepatch}),
the outer spherical surface $S_b$ with radius $r_b$, and and excised sphere $S_e$ with radius $r_e$.
The role of $S_a$ is to exclude the region near the black hole singularity. Boundary conditions must be
provided there. The role of $S_b$ is to reach the asymptotic region.  
Boundary conditions
are also imposed at $S_b$. Finaly $S_e$ is introduced to improve the angular resolution and reduce the number
of multipoles for resolving the companion object. 
The boundary value at $S_e$ is copied from the sphere of the same radius as indicated in Fig.\ref{fig:singlepatch}, 
so that the equal mass binary black holes can be calculated. 

The method that we use to solve the partial differential equations is the 
Komatsu-Eriguchi-Hachisu (KEH) method \cite{KEH} which essentially uses the representation theorem 
with a suitable chosen kernel iteratively until a fixed point is obtained. In \cite{TU2007} the KEH method
was adapted to handle multiple coordinate patches with appropriate boundary conditions.
In {\sc cocal} the construction of the
kernel is intimately related to the geometry of Fig.\ref{fig:singlepatch} and the boundary conditions
it satisfies. Denoting by $B_a$ the ball or radius $r_a$, $B_b$ the ball of radius $r_b$, and $B_e$
the ball of radius $r_e$, it is $S_a=\pd B_a$, $S_b=\pd B_b$, $S_e=\pd B_e$. Our computational domain
is $V=B_b-(B_a\cup B_e)$ and we have $\pd V=S_a\cup S_b\cup S_e$. 

A typical boundary value problem (BVP) that we encounter is
\be
\nabla^2 \Phi = S(\Phi)\quad\mbox{in }V,\qquad\qquad \mathcal{L}\Phi=f\quad\mbox{on } \pd V
\label{eq:Poisson}
\ee
where $\Phi$ can be any of the metric potentials and $\mathcal{L}$ a first order linear operator.
Following \cite[chap 3]{Jackson} we write the solution as 
\beq
\Phi(x) \,=\, \chi(x) \,+\, \Phiint(x), 
\label{eq:solver}
\eeq
where
\beqn
\Phi_{\mathrm{INT}}(x)= -\frac1{4\pi}\int_{V} G(x,x')S(x') d^{3}x' 
\qquad\qquad\qquad\nonumber \\
+ \frac{1}{4\pi} \int_{\pd V} \left[G(x,x')\na'^{a} \Phi(x')
- \Phi(x')\na'^a G(x,x') \right]dS'_a. 
\ \ \ 
\label{eq:GreenIde}
\eeqn
and $\chi(x)$ is the corresponding homegeneous solution of the BVP
\be
\nabla^2 \chi = 0\quad\mbox{in }V,\qquad\qquad \mathcal{L}\chi=f-\mathcal{L}\Phi_{\mathrm{INT}}\quad\mbox{on } \pd V
\label{eq:Poissonchi}
\ee
In Eq. (\ref{eq:GreenIde}) 
$G(x,x')$ is the flat Green's function that satisfies $\nabla^2 G(x,x') = -4\pi \dl(x-x')$.
Expanding $G$ in multipoles on a spherical coordinate system we have
\beqn
G(x,x')&=&
\frac{1}{\left|{x}-{x'}\right|}\,=\, 
\sum_{\ell=0}^\infty g_\ell(r,r') \sum_{m=0}^\ell \epsilon_m \,
\frac{(\ell-m)!}{(\ell+m)!}
\nonumber\\
&&\!\!\!\!\!\!\!
\times
P_\ell^m(\cos\theta)\,P_\ell^m(\cos\theta')
\cos m(\varphi-\varphi'), 
\eeqn
where the radial Green's function $g_\ell(r,r')$ is defined by 
\beq
g_\ell(r,r')=\frac{r_<^\ell}{r_>^{\ell+1}}, 
\eeq
with 
$
r_> := \max\{r,r'\}, \ r_< := \min\{r,r'\}, 
$
and the coefficients $\epsilon_m$ are equal to $\epsilon_0 = 1$, 
and $\epsilon_m = 2$ for $m\ge 1$.

\subsection{Implementation for Robin-Dirichlet boundary conditions}
\label{sec:RobinDirichlet}

When one computes inversion-symmetric initial data or when enforces the inner surface $S_a$
to be an apparent horizon a Robin type boundary condition for the conformal factor is obtained.
The BVP that has to be solved is
\begin{eqnarray*}
\nabla^2\Phi & = & S(\Phi)\quad\mbox{in }V  \\ 
\left[\frac{\pd\Phi}{\pd r}+\frac{\Phi}{2r}\right]_{r=r_a}& = &f  \\
\left[\Phi\right]_{r=r_b} & = & \Phi_b
\end{eqnarray*}
where $f, \Phi_b$ are known functions. The corresponding BVP for the homogeneous solution is
\begin{subequations}
\begin{align}
\nabla^2\GX & = 0\quad\mbox{in }V    \label{eq:lap_chi} \\ 
\left[\frac{\pd\GX}{\pd r}+\frac{\GX}{2r}\right]_{r=r_a}& = f -
\left[\frac{\pd\Phi_{\mathrm{INT}}}{\pd r}+\frac{\Phi_{\mathrm{INT}}}{2r}\right]_{r=r_a}  \label{eq:chi_bca}\\
\left[\GX\right]_{r=r_b} & = \Phi_b - \left[\Phi_{\mathrm{INT}}\right]_{r=r_b}   \label{eq:chi_bcb}
\end{align}
\end{subequations}
Since $r^\ell,\ r^{-\ell-1}$ are the solutions of the radial part of 
the Laplacian, we write
\beqn
\GX(x) = \frac{1}{4\pi}\sum_{\ell=0}^{\infty}\sum_{m=0}^\ell \GE_m\frac{(\ell-m)!}{(\ell+m)!}
               P_\ell^m(\cos\GJ)\times     \qquad\nonumber\\
         \left\{[A_{\ell m}r^{-\ell-1}+C_{\ell m}r^\ell]\cos(m\GP)+
                      [B_{\ell m}r^{-\ell-1}+D_{\ell m}r^\ell]\sin(m\GP)]\right\} 
\label{eq:chi}
\eeqn
where $A_{\ell m}$, $B_{\ell m}$, $C_{\ell m}$, and $D_{\ell m}$ are constants.  
From boundary conditions Eq. (\ref{eq:chi_bca}), (\ref{eq:chi_bcb}) and using
the orthogonality relations
\begin{eqnarray*}
\int_0^\pi P_\ell^m(\cos\GJ)P_{\ell'}^m(\cos\GJ)\sin\GJ d\GJ &=& 
\frac{2}{2\ell+1}\frac{(\ell+m)!}{(\ell-m)!}\GD_{\ell\ell'}  \\
\int_0^{2\pi}\sin(m\GP)\cos(m'\GP)d\GP &=& 0 \qquad\qquad\qquad\qquad\\
\int_0^{2\pi}\cos(m\GP)\cos(m'\GP)d\GP &=& \frac{2\pi}{\GE_m}\GD_{mm'}  \qquad\qquad\quad 
\end{eqnarray*}
we get
\begin{eqnarray*}
A_{\ell m}r_b^{-\ell-1}+C_{\ell m}r_b^\ell = (2\ell+1)\times \nonumber\\ 
\int_0^\pi \int_0^{2\pi} \left(\Phi_b-\Phi_{\mathrm{INT}}\right)_{r=r_b}
P_\ell^m(\cos\GJ)\cos(m\GP) d\Omega   \\[10pt]
B_{\ell m}r_b^{-\ell-1}+D_{\ell m}r_b^\ell = \frac{2(2\ell+1)}{\GE_m}\times\nonumber\\
\int_0^\pi \int_0^{2\pi} \left(\Phi_b-\Phi_{\mathrm{INT}}\right)_{r=r_b}
P_\ell^m(\cos\GJ)\sin(m\GP) d\Omega  
\end{eqnarray*}
and
\begin{eqnarray*}
-A_{\ell m}r_a^{-\ell-2}+C_{\ell m}r_a^{\ell-1} = 2\times\qquad \nonumber\\ 
\int_0^\pi \int_0^{2\pi} 
\left(f-\frac{\pd\Phi_{\mathrm{INT}}}{\pd r}-\frac{\Phi_{\mathrm{INT}}}{2r}\right)_{r=r_a}
P_\ell^m(\cos\GJ)\cos(m\GP) d\Omega   \\[10pt]
-B_{\ell m}r_a^{-\ell-2}+D_{\ell m}r_a^{\ell-1} = \frac{4}{\GE_m}\times\qquad\nonumber\\
\int_0^\pi \int_0^{2\pi} 
\left(f-\frac{\pd\Phi_{\mathrm{INT}}}{\pd r}-\frac{\Phi_{\mathrm{INT}}}{2r}\right)_{r=r_a}
P_\ell^m(\cos\GJ)\sin(m\GP) d\Omega  
\end{eqnarray*}
When we solve the above system of equations with respect to $A_{\ell m}$, $B_{\ell m}$, $C_{\ell m}$, $D_{\ell m}$
and substitute back to Eq. (\ref{eq:chi}) we get
\begin{eqnarray}
\GX(x) = \frac{1}{4\pi}\sum_{\ell=0}^{\infty}\sum_{m=0}^\ell \GE_m\frac{(\ell-m)!}{(\ell+m)!}
               P_\ell^m(\cos\GJ)\times\qquad\qquad\qquad    \nonumber\\
               \left\{ a_\ell(r)
               \int_{S_b}(\Phi_b-\Phi_{\mathrm{INT}})P_\ell^m(\cos\GJ')\cos[m(\GP-\GP')] d\Omega' \right. 
               \qquad\qquad\qquad    \nonumber \\
               \left. + b_\ell(r)
               \int_{S_a}\left(f-\frac{\pd\Phi_{\mathrm{INT}}}{\pd r}-\frac{\Phi_{\mathrm{INT}}}{2r}\right)
                         P_\ell^m(\cos\GJ')\cos[m(\GP-\GP')] d\Omega'  
               \right\}. \label{eq:chiioND}
\end{eqnarray}
where
\begin{eqnarray*}
a_\ell(r) &=& (2\ell+1)\left(\frac{r_a}{r_b}\right)^\ell
               \frac{\left(\frac{r}{r_a}\right)^\ell+\left(\frac{r_a}{r}\right)^{\ell+1}}
                    {1+\left(\frac{r_a}{r_b}\right)^{2\ell+1}}, \\[10pt]
b_\ell(r) &=& -2\frac{r_a^{\ell+2}}{r_b^{\ell+1}}
               \frac{\left(\frac{r_b}{r}\right)^{\ell+1} - \left(\frac{r}{r_b}\right)^\ell}
                    {1+\left(\frac{r_a}{r_b}\right)^{2\ell+1}}
\end{eqnarray*}
The final solution will be obtained from the iteration of
\[  \Phi(x)\ =\ \GX(x)\ +\ \Phi_{\mathrm{INT}}(x)  \]
where $\Phi_{\mathrm{INT}}$ is given by Eq. (\ref{eq:GreenIde}).

\section{Binary black hole circular orbits}
\label{sec:BBHcirc}

To solve for BBH, Eq. (\ref{eq:HC}),(\ref{eq:MC}),(\ref{eq:Kdot})
are supplemented with boundary conditions at infinity as
\begin{equation}
\left.\GC\right|_{r\rightarrow \infty} =1.0,  \qquad
\left.\GB^i\right|_{r\rightarrow \infty} =0.0,  \qquad
\left.\GA\right|_{r\rightarrow \infty} =1.0  
\label{eq:Initial_bc}
\end{equation}
so as flat space time is acquired, and at the black hole excision surface $S_a$ 
\cite{CP,KADATH} with
\begin{subequations}
\begin{align}
\left.\frac{\pd\GC}{\pd r}+\frac{\GC}{2r}\right|_{r=r_a} 
& = -\frac{\GC^3}{4}K_{ij}s^i s^j ,  \label{eq:AHpsiBC} \\
\left.\GB^i\right|_{r=r_a} & = \frac{n_0}{\GC^2}s^i + \Omega\, M^i 
 - \Omega_{\rm s}\,\GP_{\rm s}^i ,\label{eq:AHbetaBC} \\ 
\left.\GA\right|_{r=r_a} & = n_0 .   \label{eq:AHalphBC}
\end{align}
\end{subequations}
Instead of the infinity, we impose the boundary condition (\ref{eq:Initial_bc}) 
at the sphere $S_b$ with the radius $r=r_b$ in the asymptotic region.  
The boundary condition 
Eq. (\ref{eq:AHalphBC}) encodes the freedom to choose the initial slice. For BBH in quasiequilibrium
the choice for the lapse is largely irrelevant. We are taking $n_0=0.1$ on $S_a$.
In Eq. (\ref{eq:AHpsiBC}) $s^i$ is the unit normal to the sphere $S_a$. This equation enforces the 
sphere $S_a$ to be an apparent horizon. Finally Eq. (\ref{eq:AHbetaBC}) ensures that the spheres
are in equilibrium and also informs about the state of rotation of the BH. $M^i$ is 
the constant translational vector defined by $(0,d,0)$, where $d$ is the coordinate distance 
between the center of mass of the binary system and the center of the black hole excision surface. 
$\GP_{\rm s}^i$ is the rotational vector with respect to the black hole. 
Parameter $\Omega$ corresponds
to the orbital velocity of the system, and parameter $\Omega_s$ to the local rotation rate of the BH
that relates to the spin of the BH. For any values of $\Omega,\Omega_s$ a solution is obtained and
we have to choose those that correspond to circular orbits. 
In this preliminary work, parameter $\Omega_s$ is set to zero, which approximately corresponds to 
irrotational (non spinning) BBH solution. 
As it is discussed in \cite{CCGP} the 
value of $\Omega_s$ that 
leads to non spinning binaries can be found by solving Eq. (\ref{eq:HC}),(\ref{eq:MC}),(\ref{eq:Kdot})
iteratively until the quasi-local spin 
\begin{equation}
S_i = \frac{1}{8\pi}\int_{S_{a_i}} \GC^6 {\tilde A}_{jk}\GP_{s(i)}^j dS^k,     \label{eq:qls}
\end{equation}
vanishes. We will present the result with such adjustment of the spin in our forthcoming paper.

Following \cite{GGBa} the 
value of the orbital velocity $\Omega$ for a circular orbit 
is obtained by requiring equality
of the ADM mass and the Komar mass. In the conformally flat spacetime these are calculated from
\begin{eqnarray}
M_{\rm ADM} & = & -\frac{1}{2\pi}\int_{\infty}\pd_i\GC dS^i,  \label{eq:ADMmass} \\
M_{\rm Komar} & = & \frac{1}{4\pi}\int_{\infty}\pd_i\GA dS^i.  \label{eq:Komarmass}  
\end{eqnarray}
In our code the surface integrals are performed over the sphere $S_b$. 
Alternatively we can convert the surface integrals at infinity to volume integrals and surface
integrals on the BH thus providing a consistency check to the accuracy of our solution. Using
Eq. (\ref{eq:HC}) we get for the ADM mass also
\begin{equation}
M_{\rm ADM}=\frac{1}{16\pi}\int_V \GC^5{\tilde A}_{ij}{\tilde A}^{ij}dV
-\frac{1}{2\pi}\int_{S_e}\pd_i\GC dS^i -\frac{1}{2\pi}\int_{S_a}\pd_i\GC dS^i.  \label{eq:ADMvol} 
\end{equation}
The ADM mass given by Eq. (\ref{eq:ADMvol}) will be used to test the accuracy of solving
the Hamiltonian constraint.
 
The total angular momentum in a hypersurface $\Sigma_t$ is defined as
\begin{equation}
J=\frac{1}{8\pi}\int_{\infty}(K_j^i-K\GD^i_j)\GP_{\rm cm}^j dS_i = 
\frac{1}{8\pi}\int_{\infty} {\tilde A}^i_j \GP_{\rm cm}^j dS_i.  \label{eq:Jinf}
\end{equation}
Again this integral in our code is taken over the sphere $S_b$. $\GP_{\rm cm}^j$ is the rotational
vector with respect to the center of mass.
Similarly we convert the integral at infinity to an integral at the throat plus a volume integral. The 
latter vanishes due to the momentum constraint Eq. (\ref{eq:MC}) and the fact that $\GP_{\rm cm}^j$ 
is a Killing vector for the flat metric. Finally we have
\begin{equation}
J=-\frac{1}{8\pi}\int_{S_{a_1}} \GC^6 {\tilde A}^i_j \GP_{\rm cm}^j dS_i
  -\frac{1}{8\pi}\int_{S_{a_2}} \GC^6 {\tilde A}^i_j \GP_{\rm cm}^j dS_i.   \label{eq:Jthr}
\end{equation}
Comparing Eq. (\ref{eq:Jthr}) with Eq. (\ref{eq:Jinf}) we get an estimate for the violation of 
the momentum constraint. As a final check of the overall computation we consider the Smarr formula 
\begin{equation}
M_{\rm ADM} - 2\Omega J = -\frac{1}{4\pi}\int_{S_{a_1}}\GC^2\pd_i\GA dS^i
                          -\frac{1}{4\pi}\int_{S_{a_2}}\GC^2\pd_i\GA dS^i,   \label{eq:Smarr}
\end{equation}
that relates the ADM mass, the angular momentum, and the computed orbital velocity. Typically by
inserting $M_{\rm ADM}$ from Eq. (\ref{eq:ADMmass}) and the calculated orbital velocity we can
get a third value for the angular momentum of the system that can be compared against Eq. (\ref{eq:Jthr}),
and Eq. (\ref{eq:Jinf}).

Another important quantity is the irreducible mass and the binding energy of the system. 
It is $M_{\rm irr}=m_1+m_2$ with $m_i=\sqrt{A_i/16\pi}$ and 
\begin{equation}
A_i=\int_{S_{a_i}} \GC^4 dS.\    \label{eq:area}
\end{equation}
The binding energy is then $E_b=M_{\rm ADM}-M_{\rm irr}$. 

Using the H3 resolution in Table \ref{tab:BBHtest_grids}, 
a sequence of equal mass black holes is obtained whose main characteristics
(separation parameter, angular velocity, ADM mass, binding energy, and angular momentum)
are shown in Table \ref{tab:BBHseq}. The renormalization was done using the irreducible mass.
For the BH coordinate separation we used $d_s=2.5$.

\begin{figure}
\begin{center}
\includegraphics[height=70mm]{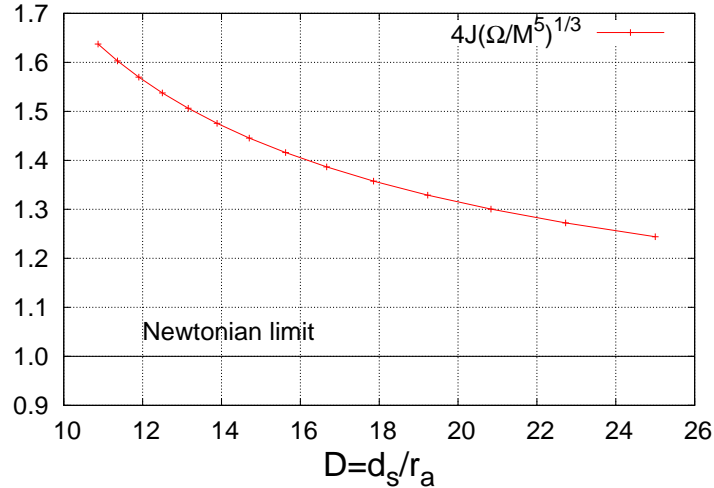}
\caption{The quantity $4J(\Omega/M^5)^{1/3}$ is plotted against the separation parameter D.}
\label{fig:plot_dev_from_3rd}
\end{center}
\end{figure}

For two point particles of individual mass $m$ moving in circular orbit of radius $R$, Kepler's
third law gives $\Omega^2=2m/(2R)^3$, and since the total angular momentum of the system is
$J=2m\Omega R^2$, we have that in Newtonian mechanics the total mass of the system $M=2m$, the
total angular momentum $J$, and the angular velocity $\Omega$ satisfy
\be
 4J\left(\frac{\Omega}{M^5}\right)^{\frac{1}{3}}=1\ .   \label{eq:JOmegaM} 
\ee
The deviation from the Newtonian value of the quantity on the left side of Eq. (\ref{eq:JOmegaM})
can be seen in Fig. \ref{fig:plot_dev_from_3rd}. The closer the black holes are, the larger the
difference between the Newtonian and general relativistic prediction.
 
\begin{figure}
\begin{center}
\includegraphics[height=70mm]{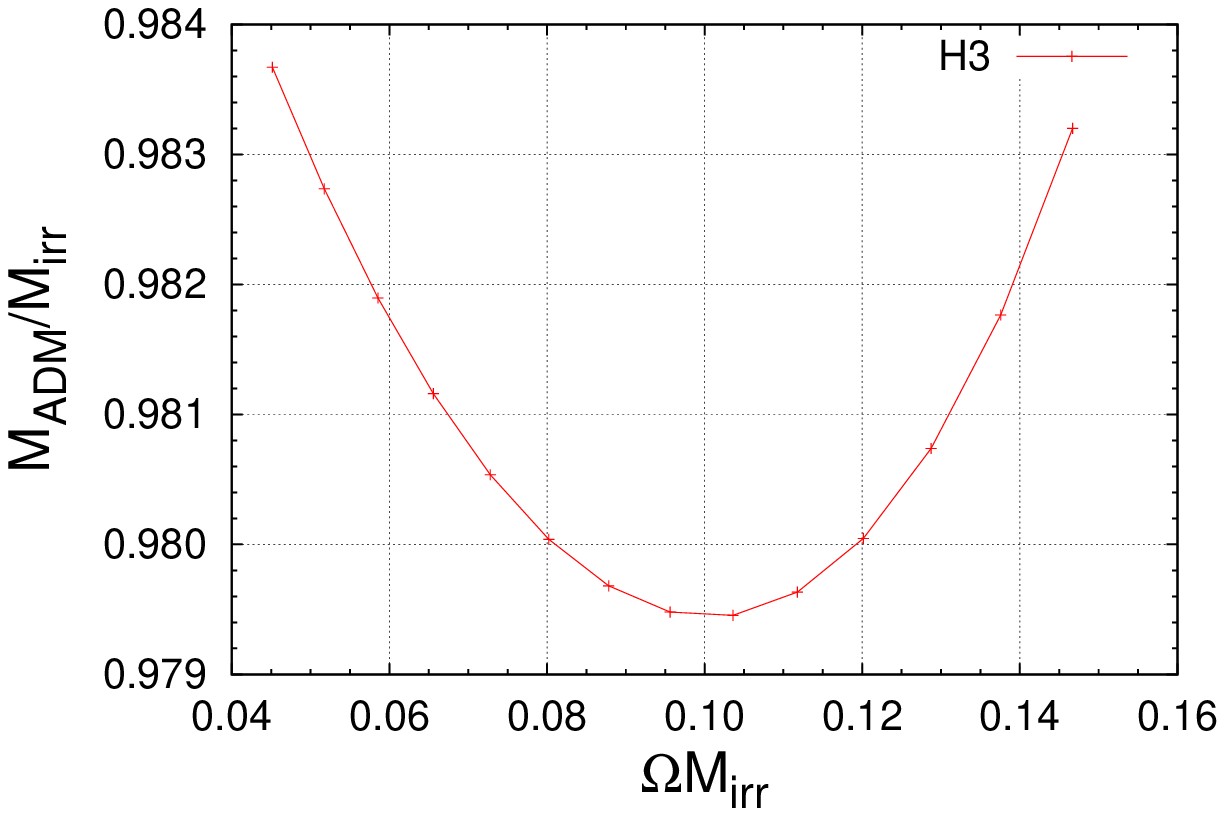}
\includegraphics[height=70mm]{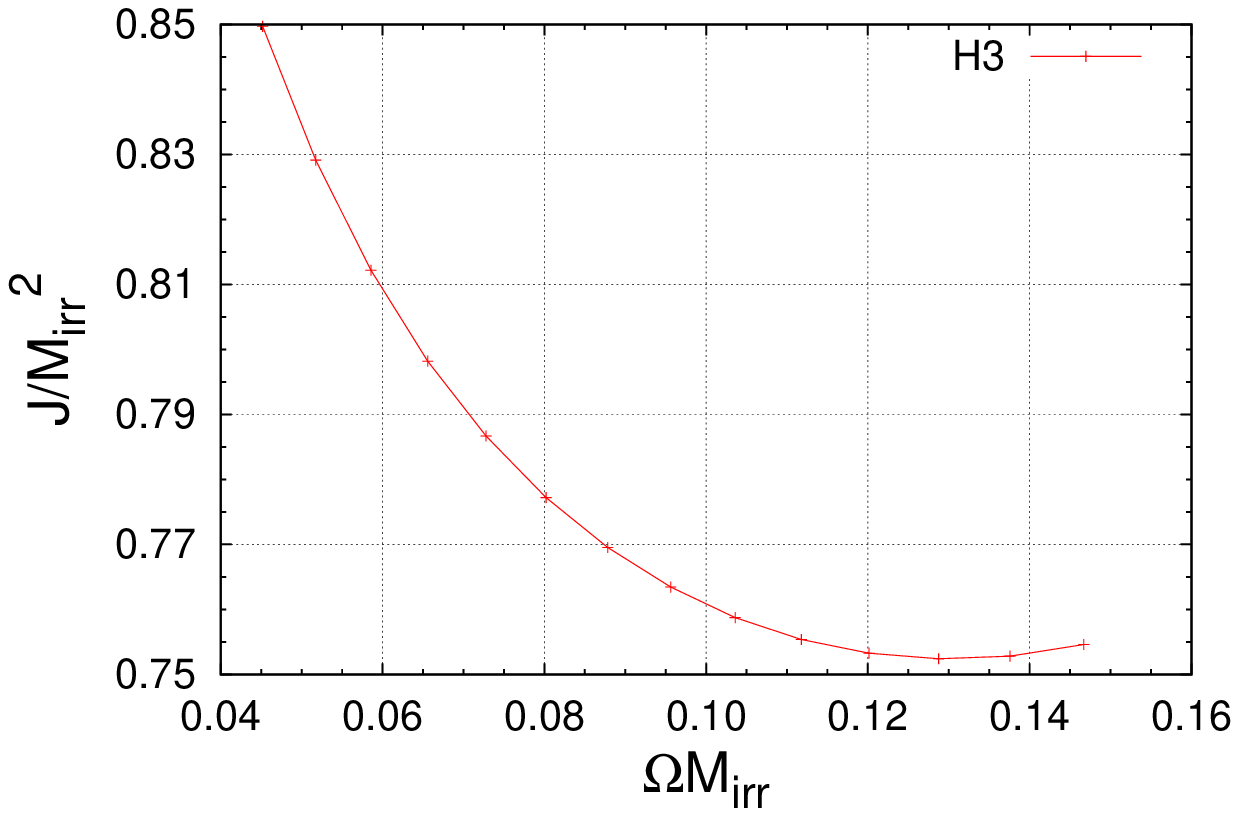}
\includegraphics[height=70mm]{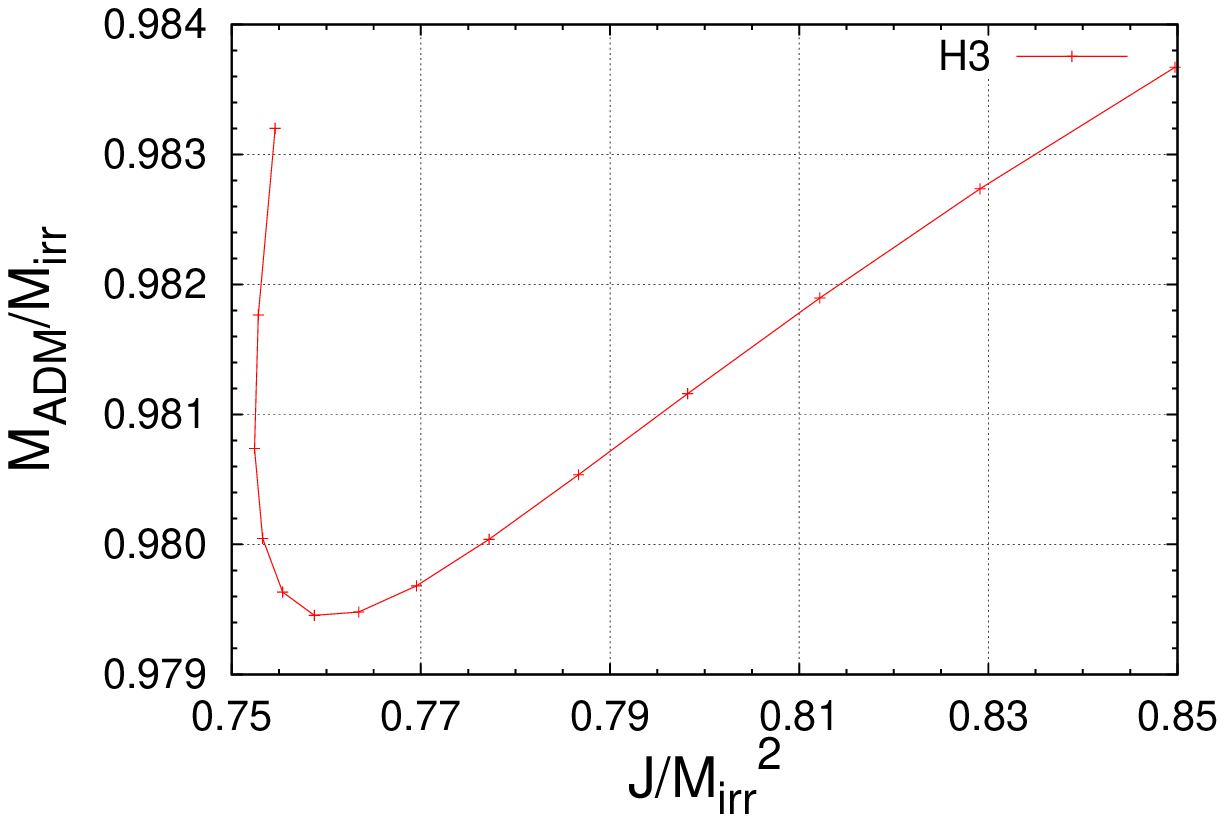}
\caption{Top panel: plot of ADM mass versus the orbital velocity. Middle panel: plot of
angular momentum versus the orbital velocity. Bottom panel: plot of ADM mass versus angular momemtum.}
\label{fig:plot_r_grid_H3}
\end{center}
\end{figure}

Plots of the ADM mass and the angular momentum versus the angular velocity can be seen in 
Fig. \ref{fig:plot_r_grid_H3}. As the black holes come together the mass and the angular momentum
exhibits a minimum that signifies the innermost stable circular orbit (ISCO). As we observe from
the top panel of Fig. \ref{fig:plot_r_grid_H3} the minimum of the ADM mass is different from the
minimum of the angular momentum (middle panel). The reason for this discrepancy is that the magnitude
of the spin as computed by Eq. (\ref{eq:qls}) is not exactly zero and therefore the sequence is not
strictly speaking irrotational. This is the reason that the characteristic cusp in the
mass versus angular momentum plot is absent (bottom panel). As we discussed above this issue is 
resolved when we iterate over $\Omega_s$ so as to make the spin, Eq. (\ref{eq:qls}), to be zero.

\begin{table}
\begin{tabular}{cccccrrrrrrr}
\hline
Type & $r_a$ & $r_b$ & $r_e$ &  $d$ &  $d_s$ & $N_r$ & $N_\theta$ & $N_\phi$ & $L$ \\
\hline
H3 & 0.1-0.23 & $10^6$ & 1.125 & 1.25 & 2.5 & 384  & 96  & 96 &  12 \\
\hline
\end{tabular}
\caption{Grid parameters used in the computation of equal mass BBH sequence.  
$(N_r,N_\theta,N_\phi)$ are the numbers of grid points in $(r,\theta,\phi)$ 
coordinates, and $L$ is the number of multipoles included in the Green function.
We vary the radius $r_a$ in the range shown in the table to compute 
the solution sequence from larger to smaller separations. }
\label{tab:BBHtest_grids}
\end{table}

\begin{table}
\begin{tabular}{cccccccccccc}
\hline
 & & & $D$ & $\Omega M_{\rm irr}$ & $M_{\rm ADM}/M_{\rm irr}$ & 
 $E_b/M_{\rm irr}$ & $J/M_{\rm irr}^2$ & & &  \\
\hline
  & & & 25.00 &  0.04516 &  0.98367 &  -0.01633 &  0.84973  & & & \\
  & & & 22.73 &  0.05175 &  0.98274 &  -0.01726 &  0.82913  & & & \\
  & & & 20.83 &  0.05856 &  0.98190 &  -0.01810 &  0.81218  & & & \\
  & & & 19.23 &  0.06558 &  0.98116 &  -0.01884 &  0.79820  & & & \\
  & & & 17.86 &  0.07281 &  0.98054 &  -0.01946 &  0.78667  & & & \\
  & & & 16.67 &  0.08023 &  0.98004 &  -0.01996 &  0.77721  & & & \\
  & & & 15.63 &  0.08783 &  0.97968 &  -0.02032 &  0.76953  & & & \\
  & & & 14.71 &  0.09563 &  0.97948 &  -0.02052 &  0.76343  & & & \\
  & & & 13.89 &  0.10360 &  0.97945 &  -0.02055 &  0.75874  & & & \\
  & & & 13.16 &  0.11178 &  0.97963 &  -0.02037 &  0.75538  & & & \\
  & & & 12.50 &  0.12015 &  0.98005 &  -0.01995 &  0.75327  & & & \\
  & & & 11.90 &  0.12874 &  0.98074 &  -0.01926 &  0.75240  & & & \\
  & & & 11.36 &  0.13757 &  0.98176 &  -0.01824 &  0.75281  & & & \\
  & & & 10.87 &  0.14669 &  0.98320 &  -0.01680 &  0.75460  & & & \\
\hline
\end{tabular}
\caption{Sequence of equal mass black holes on a maximal slice.
The normalization is done using the irreducible mass.}
\label{tab:BBHseq}
\end{table}

\section{Discussion}

We have successfully computed a sequence of conformally flat 
initial data for non-spinning equal mass BBH solutions in circular orbits.  
Several authors have calculated sequences of this kind 
as models of BBH inspiral due to the graviational wave radiation 
\cite{GGBb,CCGP,KADATH}. 
They used spectral methods in their computations, and produced numerical 
solutions in higher precision compared to ours. 
In the {\sc cocal} code, we use standard, mostly second order, finite 
difference scheme. Our method is much simpler than the spectral method, 
and hence it is easier to extend our code to include magnetic fields 
or neutron stars. Also we have demonstrated that the solutions are 
accurate enough to reproduce the results of \cite{GGBb,CCGP,KADATH}. 
Further details of the {\sc cocal} code will be discussed elsewhere.

\acknowledgements
The authors wish to thank members of the Observatory of Meudon (LUTH) for their warm hospitality
and valuable discussions. This paper is dedicated to Peter Leach on the occasion of his seventieth birthday.

\end{document}

%% file: kigou.tex
\newcommand{\beq}{\begin{equation}} 
\newcommand{\eeq}{\end{equation}} 
\newcommand{\beqn}{\begin{eqnarray}} 
\newcommand{\eeqn}{\end{eqnarray}} 
\newcommand{\pa}{\partial}
\newcommand{\na}{\nabla}

\newcommand{\zD}{{\raise1.0ex\hbox{${}^{\ \circ}$}}\!\!\!\!\!D}
\newcommand{\alone}{{\raise0.5ex\hbox{${}^{\ 1}$}}\!\!\!\!\alpha}

\newcommand{\dl}{\delta}

\newcommand{\nalam}{\mathrel{\raise0.9ex\hbox{$^\lambda$}\mkern-14mu
\lower0.0ex\hbox{$\nabla$}}}

\newcommand{\Phiint}{{\Phi_{\rm INT}}}

\newcommand{\zeroD}{{\raise1.0ex\hbox{${}^{\ \circ}$}}\!\!\!\!\!D}
\newcommand{\Lap}{\Delta}
\newcommand{\zLap}{{\raise1.0ex\hbox{${}^{\ \circ}$}}\!\!\!\!\Delta}
\newcommand{\zna}{{\raise1.0ex\hbox{${}^{\ \circ}$}}\!\!\!\!\!\nabla}
\newcommand{\zS}{{\raise1.0ex\hbox{${}^{\ \circ}$}}\!\!\!\!\!S}

